# Imaging Polarity in Two Dimensional Materials by Breaking Friedel's Law

Pratiti Deb[1*], Michael C. Cao[1], Yimo Han[1], Megan E. Holtz[1], Saien Xie[1,2], Jiwoong Park[2], Robert Hovden[1†], and David A. Muller[1,3]

[1.] School of Applied and Engineering Physics, Cornell University, Ithaca, USA.

[2.] Department of Chemistry, Pritzker School of Molecular Engineering, and James Franck Institute, University of Chicago, Chicago, IL 60637, USA

[3.] Kavli Institute at Cornell for Nanoscale Science, Ithaca, USA

Corresponding Author:

David A. Muller
227 Clark Hall
Ithaca, NY, 14853-2501
dm24@cornell.edu
t. 607-255-4065
f. 607-255-7658

Declarations of interest: None

---

[*] Current address: Pritzker School of Molecular Engineering, University of Chicago, Chicago, IL 60637, USA
[†] Current address: Department of Materials Science and Engineering, University of Michigan, Ann Arbor, MI 48109

**Abstract**

Friedel's law guarantees an inversion-symmetric diffraction pattern for thin, light materials where a kinematic approximation or a single-scattering model holds. Typically, breaking Friedel symmetry is ascribed to multiple scattering events within thick, non-centrosymmetric crystals. However, two-dimensional (2D) materials such as a single monolayer of $MoS_2$ can also violate Friedel's law, with unexpected contrast between conjugate Bragg peaks. We show analytically that retaining higher order terms in the power series expansion of the scattered wavefunction can describe the anomalous contrast between $hkl$ and $\overline{hkl}$ peaks that occurs in 2D crystals with broken in-plane inversion symmetry. These higher-order terms describe multiple scattering paths starting from the same atom in an atomically thin material. Furthermore, 2D materials containing heavy elements, such as $WS_2$, always act as strong phase objects, violating Friedel's law no matter how high the energy of the incident electron beam. Experimentally, this understanding can enhance diffraction-based techniques to provide rapid imaging of polarity, twin domains, in-plane rotations, or other polar textures in 2D materials.

**Keywords:**

**2D Materials, Polar, Non-centrosymmetric, Diffraction, Pixelated detector**

**Highlights:**

- **Diffraction patterns from polar 2D materials break Friedel's law.**
- **Monolayer transition metal dichalcogenides are strong phase objects.**
- **Diffraction asymmetry is traced to higher-order scattering, even in monolayers.**
- **Pixelated detectors enable rapid fully quantitative measure of polarity in 2D materials.**

# 1. Introduction

Atomically thin two-dimensional (2D) materials are usually interpreted in terms of kinematic scattering and treated in the weak phase approximation (WPA), leading us to expect a symmetric diffraction pattern[1-4]. However, dark field transmission electron microscopy (DF-TEM) on two-dimensional molybdenum disulfide ($MoS_2$) has shown a difference in the intensities at the $hkl$ and $\overline{hkl}$ peaks of the diffraction pattern[5]. This anomalous contrast between the conjugate peaks can be traced to the breaking of in-plane inversion symmetry, or non-zero polarity, of these materials. The polarity can be used to rapidly image the twin domains, or adjacent crystals with an in-plane rotation of 180º, in 2D transitional metal dichalcogenide (TMD) crystals. TMDs are a class of materials of composition $MX_2$ where M is a transition metal bonded to chalcogens, X, in the planes above and below. In its monolayer form, the $2H-MX_2$ polytype has a direct band gap arising from broken inversion symmetry, making it a promising candidate for optoelectronic applications[6-8].

In crystallography, Friedel's law states that a diffraction pattern will have equal intensities for both the $hkl$ and $\overline{hkl}$ conjugate diffracted beams: I ($hkl$) = I ($\overline{hkl}$), as expected for kinematic scattering from a real potential[9, 10]. Historically, the violation of Friedel's law was attributed to multiple scattering from different atoms in a thick crystal, in a series of weak scattering events [11-15]. It is thus surprising that similar effects are seen in monolayer materials. In 2D materials, tilting of the specimen has been used to break Friedel symmetry and reveal stacking order[16].

Here, we demonstrate that a single monolayer of 2D TMD materials even without any tilt can break Friedel's law. Even single-atom-thick layers of a 2D material approximated by a zero-thickness film can behave as strong scatterers, as they introduce terms beyond the weak phase and weak amplitude approximations. This strong scattering behavior results in anomalous contrast in the diffraction patterns that reflects the real space polarity of the crystal. In thick crystals, these effects can be described by Bloch-wave theory, and we show that similar behavior can be recovered even for 2D materials using only the strong phase approximation.

Recent advances in direct-electron detectors for electron diffraction have provided new opportunities for studying localized phenomena in crystal structures[17-20]. Our newly-developed high dynamic range electron microscopy pixel array detector (EMPAD)[21] can easily acquire unsaturated diffraction patterns and thus gives a fully quantitative measure of polarity in these monolayer 2D materials. Figure 1a shows a map of polarity for a lateral heterojunction sample of tungsten disulfide (2H-$WS_2$) / tungsten diselenide (2H-$WSe_2$), while Figure 1(b-c) show individual diffraction patterns from the 2H-$WS_2$ and 2H-$WSe_2$ regions respectively, acquired using the EMPAD. We observe that Friedel's law is broken in the first order rings for diffraction spots on opposite sides of the central beam. We define the anomalous contrast that describes the deviation from Friedel's law, and reflects the degree of polarity, as $P = [I(hkl) - I(\overline{hkl})]/[I(hkl) + I(\overline{hkl})]$.

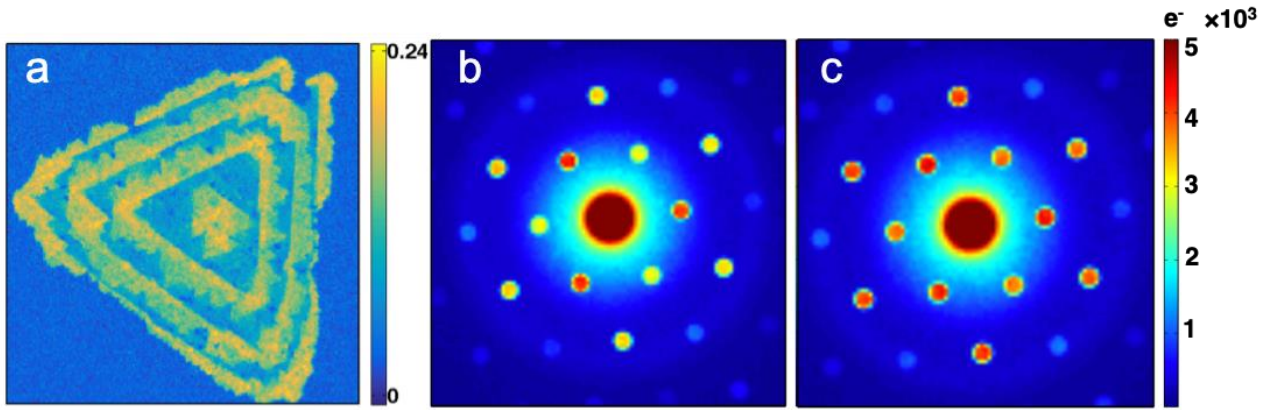

**Figure 1**. **a**, Map of polarity of $WS_2/WSe_2$ lateral heterojunctions acquired using an electron microscope pixel array detector at a beam energy 120 keV. Polarity from the anomalous contrast is calculated as $[I(010) - I(0\bar{1}0)]/[I(010) + I(0\bar{1}0)]$. **b**, A diffraction pattern from the $WS_2$ region (yellow in panel a) demonstrates non-centrosymmetry arising from the polarity, visible as an asymmetry in intensities between conjugate first-order diffraction peaks. **c**, The polarity is more difficult to observe in the diffraction pattern from $WSe_2$ region (light blue in panel a), where there is a smaller difference in atomic numbers than in $WS_2$.

The anomalous contrast is more pronounced in the WS$_2$ region than the WSe$_2$ region, due to the larger difference in atomic numbers of the elements in WS$_2$ than in WSe$_2$.

## 2. Theory

As stated above, Friedel's law requires that the diffraction intensities for the $hkl$ and $\overline{hkl}$ crystal reflections are equal, or in terms of the electron wavefunction in reciprocal space, $\Psi(\boldsymbol{q})$, that for each diffraction vector, $\boldsymbol{q}$, $|\Psi(\boldsymbol{q})|^2 = |\Psi(-\boldsymbol{q})|^2$. This law holds in the kinematic scattering limit for diffraction using the WPA or the first Born approximation. In this approximation, the diffraction intensity is directly proportional to the squared amplitude of the structure factor, $|F(\boldsymbol{q})|^2$, resulting in a centrosymmetric diffraction pattern, even when $F(\boldsymbol{q})$ and the underlying structure itself are not centrosymmetric[22]. We will therefore need to go beyond the WPA to explain the asymmetric scattering. The electron wave function in the strong phase approximation (SPA) is:

$$\Psi(\boldsymbol{r}) = \Psi_0(\boldsymbol{r}) \exp(i\sigma V(\boldsymbol{r})) , \qquad (1)$$

where $\sigma$ is the interaction parameter and $V(\boldsymbol{r})$ is the projected atomic potential of the material. The SPA is also known as the Eikonal approximation[23] and is equivalent to a Wentzel-Kramers-Brillouin (WKB) approximation for free electrons[24-27]. Here we treat 2D materials as zero-thickness films oriented perpendicular to the electron beam—i.e. interlayer scattering is not permitted and the Ewald Sphere is flat.

In the WPA, we only include the linear order term, $\sigma V(\boldsymbol{r})$, from a power series expansion of (1). The wavefunction becomes:

$$\Psi(r) = \Psi_0(r)\bigl(1 + i\sigma V(r)\bigr). \tag{2}$$

Taking the Fourier transform of this wavefunction we get

$$\Psi(q) = \frac{1}{2\pi} \int dr \exp(-iq \cdot r)\, \Psi_0(r)\bigl(1 + i\sigma V(r)\bigr), \tag{3}$$

and taking the square of its absolute value, we find the diffraction intensity

$$|\Psi(q)|^2 = \begin{aligned} &|\Psi_0(q)|^2 + \frac{\sigma e a_0}{\pi} \mathrm{Im}\{\Psi_0(q)[\Psi_0^*(q) \otimes F^*(q)]\} \\ &+ \frac{(\sigma e a_0)^2}{4\pi^2} |\Psi_0(q) \otimes F(q)|^2 + O(\sigma^2) \end{aligned}, \tag{4}$$

Here $\Psi_0(q)$ is the Fourier transform of $\Psi_0(r)$, $e$ is the charge of an electron, $a_0$ is the Bohr radius and $F(q)$ is the structure factor, defined as

$$F(q) = \frac{1}{2\pi e a_0} \int dr \exp(-iq \cdot r) V(r). \tag{5}$$

Equation (4) can be simplified further when diffraction disks do not overlap by noting that the first two terms are non-zero only for the central beam, so the contrast of the diffracted beams depend only on the last term, ie. $\frac{(\sigma e a_0)^2}{4\pi^2}|\Psi_0(q) \otimes F(q)|^2$. From the properties of the Fourier transform of a real function, we know that $F(q)$, the Fourier transform of $V(r)$ must satisfy $F(-q) = F^*(q)$. As a result, eq. (4) is symmetric between $q$ and $-q$ for the diffracted beams, and there is no anomalous contrast between the $hkl$ and $\overline{hkl}$ peaks in the electron diffraction pattern represented by the $\sigma^2$ term, since the phase information is not preserved in the pattern. This is the WPA derivation of Friedel's law.

However, retaining higher order terms in the power series expansion of the wavefunction from eq. (1) gives rise to an anomalous contrast between the $hkl$ and $\overline{hkl}$ peaks, which can be traced to the polarity or breaking of in-plane inversion symmetry of the crystal. Retaining higher order terms in the power series expansion of the wavefunction, we get:

$$\Psi(r) = \Psi_0(r)\left(1 + i\sigma V(r) - \frac{\sigma^2 V^2(r)}{2} - \frac{i\sigma^3 V^3(r)}{6} + \cdots\right). \tag{6}$$

In order to get the diffraction intensity $|\Psi(q)|^2$, we Fourier transform this third order expansion and take the square of its absolute value:

$$|\Psi(q)|^2 = \begin{aligned}&|\Psi_0(q)|^2 + \frac{\sigma e a_0}{\pi}\text{Im}\{\Psi_0(q)[\Psi_0^*(q) \otimes F^*(q)]\} \\ &+ \frac{(\sigma e a_0)^2}{4\pi^2}\{|\Psi_0(q) \otimes F(q)|^2 - \text{Re}[\Psi_0(q)(\Psi_0^*(q) \otimes F_2^*(q))]\} \\ &+ \frac{(\sigma e a_0)^3}{8\pi^3}\left\{\begin{array}{l}\text{Im}[(\Psi_0(q) \otimes F(q))(\Psi_0^*(q) \otimes F_2^*(q))] \\ -\frac{1}{3}\text{Im}[\Psi_0(q)(\Psi_0^*(q) \otimes F_3^*(q))]\end{array}\right\} \\ &+ \cdots\end{aligned}, \tag{7}$$

where

$$F_2(q) = F(q) \otimes F(q), \tag{8a}$$

$$F_3(q) = F(q) \otimes F(q) \otimes F(q). \tag{8b}$$

For the weak phase approximation, the diffraction intensity only has the leading $O(\sigma^2)$ term, which is directly proportional to $|F(q)|^2$ and does not retain any phase information, giving rise to a centrosymmetric diffraction pattern. However, in the higher order expansion, one of the third order terms is antisymmetric. This is the leading-order contribution to polarity, $P_3$, labelled as such as reminder that it is a third-order correction:

$$P_3 = \frac{(\sigma e a_0)^3}{8\pi^3}\text{Im}\left[(\Psi_0(\mathbf{q}) \otimes F(\mathbf{q}))(\Psi_0^*(\mathbf{q}) \otimes F_2^*(\mathbf{q}))\right]. \tag{9}$$

To simplify this further, we use a general structure factor for a periodic potential that is written as

$$F(q) = \sum_G U_G \exp(i\phi_G)\,\delta(\mathbf{k} - \mathbf{G})\,, \tag{10}$$

where $\mathbf{G}$ is a reciprocal lattice vector, $U_G$ is the Fourier coefficient and $\phi_G$ is the phase factor. Using this potential and assuming non-overlapping disks, eq. (9) can be simplified to

$$P_3 = \frac{(\sigma e a_0)^3}{8\pi^3} \sum_{G_1, G_2} |\Psi_0(q - G_1)|^2 U_{G_1} U_{G_2} U_{G_1 - G_2} \sin(\phi_{G_1} - \phi_{G_2} - \phi_{G_1 - G_2}), \quad (11)$$

In symmetric or non-polar materials, $V(r) = V(-r) \therefore F(q) = F(-q)$. Furthermore, since for real potentials, $F(-q) = F^*(q)$, $F(q)$, which is the Fourier transform given by eq. (5), must be pure real. This sets all the phases $\phi_G$ to zero or $\pi$ and makes eq. (11) equal to zero for all non-polar materials, resulting in a centrosymmetric diffraction pattern.

The case for polar crystals is more interesting. In fact, some of the terms in eq. (11) are familiar from three-beam diffraction theory for non-centrosymmetric bulk crystals (see section 5.6.3 of Zuo and Spence[28]). In particular, the phase term in eq. (11) is known as a three-phase invariant, $\Psi = \phi_{G_1} - \phi_{G_2} - \phi_{G_1 - G_2}$, because it is invariant under a change in origin of $V(r)$. Thus, eq. (11) remains zero for non-polar materials even under a shift in origin, ensuring a centrosymmetric diffraction pattern. The invariance of $\Psi$ also implies that the sign of polarity for polar materials is independent of the choice of origin. For polar materials, $\sin(\Psi)$ can serve as a good order parameter for describing a component of the polarity.

Physically, the higher order terms in eq. (7-9) can be thought of as multiple scattering paths starting from the same atom. Each convolution adds an additional family of scattering events that constructively or destructively interfere to modify the phase factors and give rise to an asymmetry in the diffraction peaks, as illustrated in Figure 2(b-c). Looking at the intensity of a particular diffraction spot at $\vec{G}_1$, we can remove one summation from eq. (11) to get

$$P_3(G_1) = \frac{(\sigma e a_0)^3}{8\pi^3} \sum_{G_2} |\Psi_0(q - G_1)|^2 U_{G_1} U_{G_2} U_{G_1 - G_2} \sin(\Psi). \quad (12)$$

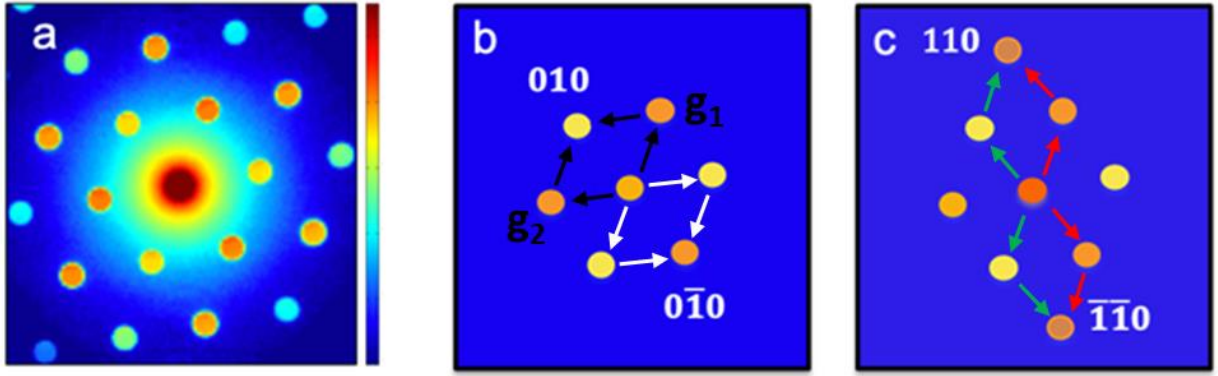

**Figure 2. a.** Diffraction pattern acquired from monolayer WS$_2$ on the EMPAD at 60 keV clearly shows the polarity, i.e. the asymmetry between the *hkl* and $\overline{hkl}$ peaks, for first order and third order ring of diffraction peaks, while peaks in the second ring are symmetric. **b-c,** The asymmetry arises from the higher order terms in the power series expansion of the electron wavefunction. This is a cartoon representation of the physical interpretation of the higher order terms as multiple scattering paths from the same atom in reciprocal space, modifying the diffraction peak intensities through constructive or destructive interference. **b** shows a second order scattering event, where the black scattering paths are constructive, so add in phase (orange/orange) to the first order spot 010. The white paths for the conjugate spot $0\overline{1}0$ also add in phase but accumulate a different net phase (yellow/yellow) leading to a different sin(Ψ) and hence a different net intensity. **c** shows second order scattering paths, where the red and green scattering paths accumulate opposite phases (orange/yellow) for spot 110. The same paths and cancellations are found for the conjugate spot $\overline{1}\overline{1}0$, leading to a symmetric pair of diffraction intensities.

The diffraction intensity depends on a sum over all closed paths including the incident beam and two scattering events, $G_1$ and $G_2$, which accumulates a phase shift sin(Ψ). We can see in Figure 2a that the first order diffraction spots break symmetry from this phase accumulation, while the second order spots do not. In Figure 2b, we can see graphically that for a given first order diffraction spot 010, there are two contributing scattering pathways that pass through two spots of dimmer intensity. For the spot $0\overline{1}0$, the two contributing scattering pathways go through two spots of brighter intensity, leading to the symmetry between 010 and $0\overline{1}0$ being broken. For the 2D hexagonal lattice, characteristic of the materials investigated here, further analytic simplification of eqn (12) is performed in Appendix A. For the first diffraction ring, as spots $\boldsymbol{g_1}, \boldsymbol{g_2}$ and $\boldsymbol{g_1 + g_2}$ all lie at the same radial distance, their amplitudes U are all the same, and their phases are simply related – they are either $\phi_{G_1}$ or $-\phi_{G_1}$ (orange or yellow). For the spots of interest in Figure 2b, $\pm(\mathbf{g_1} + \mathbf{g_2})$, we find the three-phase invariant simplifies to $\Psi = \pm 3\phi_{G_1}$, and the leading order contribution to the polarity becomes $P_3(\mathbf{g_1} + \mathbf{g_2}) \propto U_1^3 \sin(3\phi_{g_1})$, and is equal and opposite for its Friedel pair at $-(\mathbf{g_1} + \mathbf{g_2})$.

To understand the atomic-number dependence of the degree of polarity, we express the leading-order polarity correction for the 1st ring of diffraction peaks, $P_3(\mathbf{010})$ in terms of the atomic-form factors:

$$P_3(\mathbf{010}) = \frac{3\sigma^3\sqrt{3}}{64\pi^6} |\Psi_0(\mathbf{q} - \mathbf{010})|^2 f_{a1} f_{b1} (f_{a1} - f_{b1}), \tag{13}$$

where $f_{a1}$ and $f_{b1}$ are the form factors for the A site and B site atoms at the first-order scattering angle (See appendix A). Equation 13 explicitly shows that polarity is dependent on the form factor difference between the A site and B site. (Note that for the TMD materials with formula unit AB$_2$, the form factor for the projected structure, $f_{b1}$, is double the single-atom form factor.) Graphene is appropriately non-polar while polarity of other materials scale approximately with the difference in Z number (Figure A.2 in appendix A). The precise dependence of atomic form factors on atomic numbers depends on the scattering angle, with forward scattering scaling as $\sim Z^{1/3}$ in the Thomas-Fermi model to $Z$ at very high angles [11-15]. Appendix Figure A.2 shows the dependence is closer to linear. For normal diffraction intensities we are used to seeing the measurement scaling as the square of the form factor, so the linear dependence from a 3rd order term is perhaps a bit surprising, but equation (13) makes explicit the linear origin.

A similar scattering analysis for the second ring of spots (Appendix A) leads to a zero three-phase invariant, and hence zero polarity. As shown in Figure 2c, we can see that for a given pair of conjugate second order diffraction spots 110 and $\bar{1}\bar{1}0$, the two sets of scattering pathways go through diffraction spots of the same intensity, leading to symmetric spots. The preservation of phase information (through $\sin(\Psi)$) in the closed, multiple scattering paths give rise to the asymmetry between conjugate diffraction pairs and reflects the polarity in the diffraction pattern.

From the higher order expansion of the electron wavefunction, eq. (7), we can demonstrate that even though the anomalous contrast decreases with increasing incident electron beam energy, it is non-vanishing for certain polar 2D materials no matter how high the incident beam energy. The relativistic interaction parameter $\sigma$ is defined as:

$$\sigma = \frac{2\pi}{\lambda E_0}\left(\frac{m_0 c^2 + eE_0}{2m_0 c^2 + eE_0}\right). \tag{13}$$

Here $m_0$ is the rest mass of the electron, $\lambda$ is the wavelength of the electron and $E_0$ is the energy of the incident electron beam. This interaction parameter tends to a finite value as the incident beam energy $E_0 \to \infty$. Substituting for the electron wavelength $\lambda$ in terms of beam energy, we get:

$$\lim_{E_0 \to \infty} \sigma = \lim_{E_0 \to \infty} \frac{2\pi}{\lambda E_0}\left(\frac{m_0 c^2 + eE_0}{2m_0 c^2 + eE_0}\right) = \frac{e}{\hbar c}. \tag{14}$$

Thus, for finite V, $\sigma V$ cannot be made arbitrarily small, and higher order terms in $\sigma V$ will always be significant for low-order diffraction beams in heavy elements, even at high beam energies. In Figure 3, we plot the diffraction intensities of the $010$ and $0\bar{1}0$ peaks of polar 2D materials—2H-WS$_2$ and boron nitride (BN)—as a function of the interaction parameter $\sigma$, artificially allowing $\sigma$ to go to 0, beyond the relativistic limit we demonstrated. We can see that the intensities at the two peaks for WS$_2$ only converge at a value of $\sigma$ that is in the relativistically-forbidden range, i.e. the shaded region. However, for BN, the intensities at the two peaks practically converge at a relativistically-allowed value of $\sigma$. Figure 3 thus shows that a monolayer of a polar material with heavy elements such as $WS_2$ acts as a strong phase object no matter how high the energy of the incident electron beam. The WPA fails for all beam energies in this case due to the fundamental relativistic limit of the speed of the light.

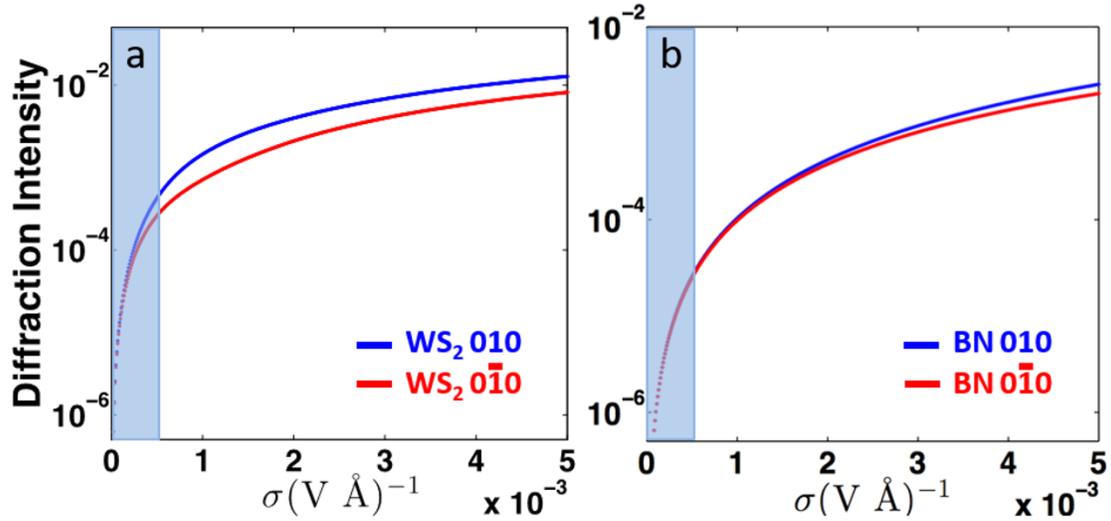

**Figure 3. a**. The diffraction intensities for the $010$ and $0\bar{1}0$ peaks for $WS_2$ are plotted as a function of the interaction parameter σ. Here, σ is artificially forced to 0, beyond its relativistic limit. We see that the intensities at the two peaks converge only at a value of σ which is in the relativistically-forbidden range, i.e. the shaded region. **b,** The diffraction intensities for the $010$ and $0\bar{1}0$ peaks for BN are plotted as a function of the interaction parameter σ. Here, unlike in **a,** the intensities of the two peaks are much closer for relativistically-allowed values of σ, consistent with the weaker scattering and smaller atomic number difference for B and N than that for W and S.

We see that multiple scattering (understood as the higher order terms in the expansion of the SPA of the electron wavefunction) can break Friedel's law within a single atomic layer (zero-thickness film) in polar 2D materials. Single layer 2H-TMDs, which contain heavy elements and mirror plane symmetry along the surface normal (c-axis) but lack in-plane inversion symmetry, provide ideal test materials.

## 3. Methods

We perform dark field transmission electron microscopy (DF-TEM) on $WS_2/WS_xSe_{2-x}/WSe_2$ lateral heterojunctions at 120 keV, as shown in Figure 4. We also perform experiments using our newly-developed high dynamic range electron microscope pixel array detector (EMPAD) to extract quantitative diffraction information from lateral heterojunctions of TMDs on nitride substrate as well as monolayer TMDs to measure their polarity. The EMPAD acquires a diffraction pattern in .86 ms at each scan position of the electron beam, and has a linear response and high dynamic range, allowing us to avoid saturation and access fully quantitative information[21]. We perform EMPAD experiments on the lateral heterojunctions of TMDs at 120 keV and on the monolayer TMDs at 60 keV on an FEI Titan Themis S/TEM, operated at 60-300 kV in scanning transmission electron microscopy (STEM) mode. The $WS_2$ and $WSe_2$ lateral heterojunctions were grown by metal-organic chemical vapor deposition (MOCVD). The individual $WS_2$ and $MoS_2$ samples were grown by MOCVD with mostly monolayer coverage on silicon dioxide ($SiO_2$), delaminated in deionized water to separate the film from the substrate and then transferred to Quantifoil copper TEM grids with 2μm holes.

We also study the anomalous contrast due to polarity in the 2D TMD materials through simulations. The exit electron wavefunction through a specimen is calculated using a multislice algorithm by sequentially propagating an incident wave through atomically-thin slices using a strong phase approximation (SPA) at each slice [11-15]. Due to instabilities in the sampling of atomic potentials projected onto a discrete grid near their origins, multislice codes that generates these potentials in real space do not reliably reproduce the polarity in the diffraction pattern. Instead, we implement code to remove the singularity by adding a finite nuclear radius. We calculate the diffraction intensity from the full electron wavefunction, i.e. the strong phase approximation (SPA). The projected crystal potential generated by the code in the SPA is a $1024 \times 1024$ matrix. We then fast Fourier transform this matrix, take its absolute value and square the absolute value to get the diffraction intensity. For a monolayer material, the SPA is equivalent to running multislice with the layer projected into a single slice – i.e. there is no propagation step inside the material. The system of units is such that the atomic potential $V(r)$ is in Volt-Angstroms ($V \cdot Å$). In addition to our multislice simulations using the SPA, we also calculate the diffraction intensities from a power series expansion of the wavefunction up to $O(\sigma^3)$ terms, i.e. including up to $O(\sigma^6)$ order terms in the diffraction intensity. In case of the power series expansion, the diffraction intensity is calculated from convolutions of the structure factor for a particular scattering vector $\boldsymbol{q}$.

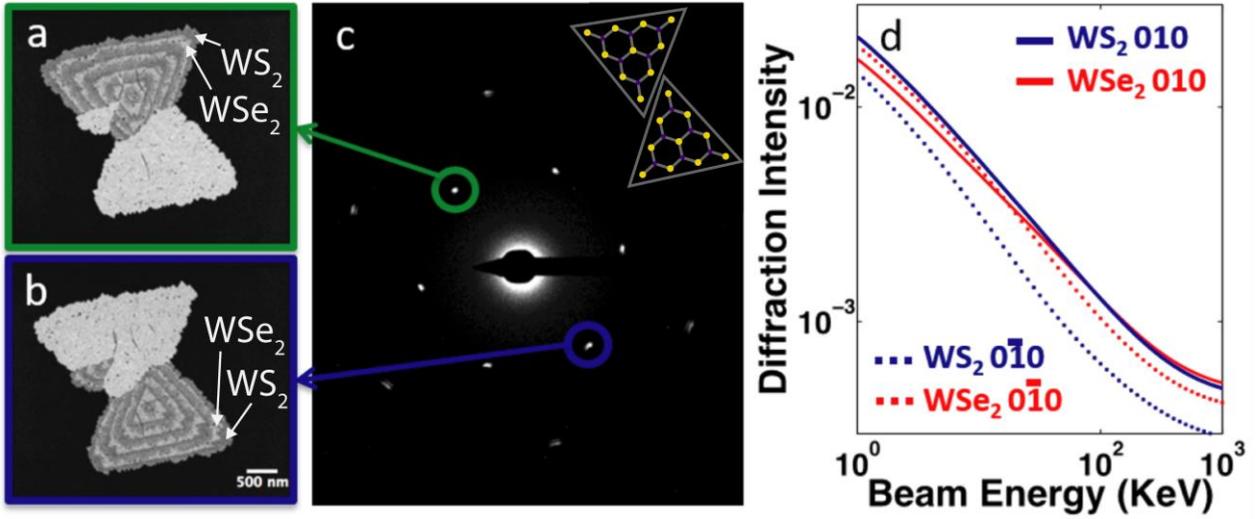

**Figure 4. a-b**, 120 keV DF-TEM images formed from 010 and 0$\bar{1}$0 peaks of the diffraction pattern of monolayer WS$_2$/WS$_x$Se$_{2-x}$/WSe$_2$ lateral heterojunctions demonstrate the polarity of the sample in the reversed contrast of the images the two dark-field images. **c**, Diffraction pattern showing the peaks corresponding to the images in **a** and **b**. The inset shows the lattice schematic of the sample in **a** and **b**, where each triangle has a similar structure to Fig. 1a due to the same growth condition. **d**, The normalized diffraction intensity for WS$_2$ and WSe$_2$ is plotted as a function of beam energy for 010 and 0$\bar{1}$0 peaks. The difference between the two peaks is greater for WS$_2$ than for WSe$_2$ correlating with the magnitude of ΔZ, where ΔZ =Z(A)-[n×Z(B)], and Z is the atomic number of A and B in the polar material AB$_n$, is greater WS$_2$ is greater than that for WSe$_2$. The difference between the 010 peaks for the two materials is also much smaller than the corresponding difference between the 0$\bar{1}$0 peaks, explaining the difference in contrast between the two triangles in **a-b**.

## 4. Results

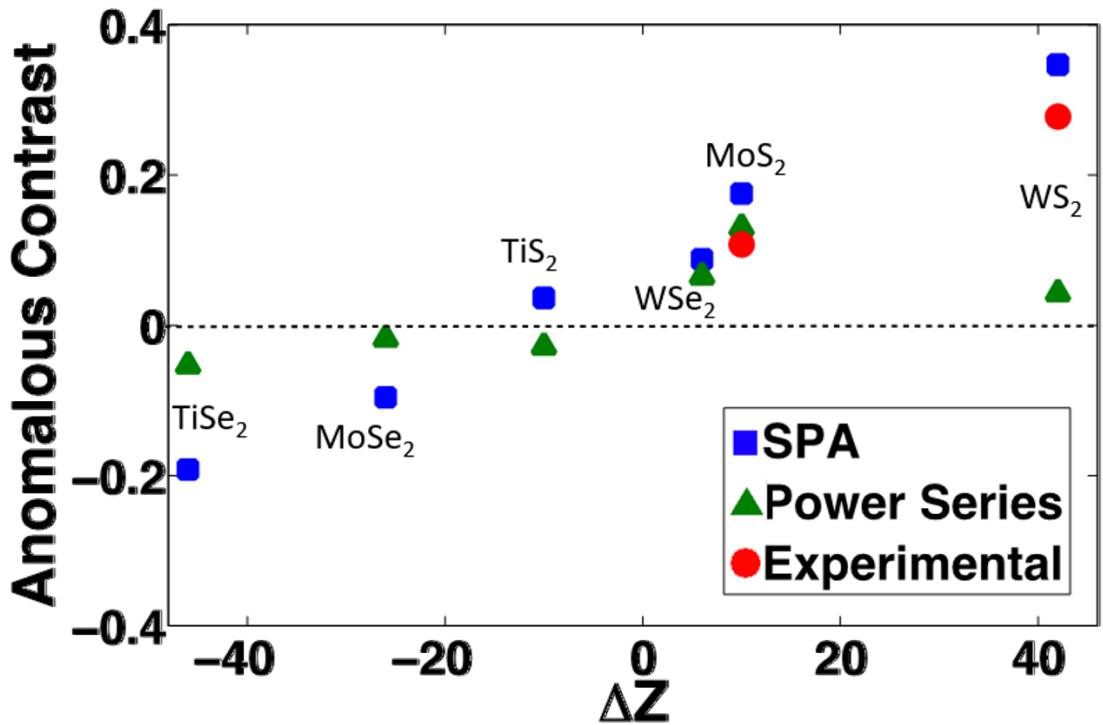

**Figure 5.** The anomalous contrast, $P = [I(hkl) - I(\overline{hkl})]/[I(hkl) + I(\overline{hkl})]$, calculated from the strong phase approximation (SPA), a power series expansion of the Born wave function, and EMPAD experimental data are plotted as a function of $\Delta Z$. We see a clear positive correlation between the contrast and $\Delta Z$ in the SPA, with the experimental data in close agreement. For the power series expansion, deviations from the trend are most apparent for the materials with the highest $\Delta Z$ (WS$_2$ and TiSe$_2$)

In Figure 4(a-c), dark field TEM images formed from Friedel pairs of diffraction peaks for a lateral heterojunction of TMDs WS$_2$ and WSe$_2$ show the polarity of the sample. The calculated diffraction intensity from our implementation of the SPA of the electron wavefunction is shown for WS$_2$ and WSe$_2$ as a function of incident electron beam energy in Figure 4d. The difference between the $010$ peak intensities (solid lines in Figure 4d) for the two materials is much lower than the corresponding difference between the $0\bar{1}0$ peak intensities (dotted lines in Figure 4d). This corroborates the reversal of contrast in the triangle heterostructures in Figure 4(a-b), which arises from the fact that the real space image in Figure 4a is mapped from one diffraction peak in Figure 4c while the image in Figure 4b is mapped from its conjugate peak.

The relative anomalous contrast, P, arising from the polarity, shows a correlation with the atomic number difference, $\Delta Z = Z(A) - n \times Z(B)$, where $Z$ is the atomic number of elements $A$ and $B$ in the polar material $AB_n$ (See Appendix A). In Figure 5, we explore the trends in anomalous contrast as a function of $\Delta Z$ for the common sulfide and selenide TMDs. The results from the EMPAD experiments performed at incident beam energy 60 keV on the monolayer $WS_2$ and $MoS_2$ samples are plotted along with the measures of polarity from the SPA and analytic calculations from the power series expansion of the electron wavefunction. For the truncated power series expansion, the anomalous contrast deviates significantly from the SPA and the experimental measure for the materials with the highest absolute value of $\Delta Z$, i.e. $WS_2$ and $TiSe_2$. This is because the phase shifts for these heavy materials with high atomic potential do not converge for the $O(\sigma^3)$ terms, which is the highest order our calculation includes, again supporting the argument that these monolayer TMD materials are acting as strong phase objects. Thermal vibrations, often approximated by an absorptive potential, may also contribute to the breakdown of Friedel symmetry. However, the expected absorptive correction, from the attenuation of the Bragg peaks is only ~1% in $MoS_2$ at the $1^{st}$-order Bragg peaks and 2-3% at the $2^{nd}$-order peaks, based on a 0.075 Å average in-plane atomic displacement [29]. This is too small to explain the 10% change seen at the $1^{st}$-order peaks in $MoS_2$,

The scaling of anomalous contrast with the degree of polarity in the material also explains the differences in polarity contrast for $WS_2$ and $WSe_2$ seen in Figures 1 and 4. The diffraction pattern from the $WS_2$ region of the sample in Figure 1b, shows a greater difference in the diffraction intensities in the 010 and $0\bar{1}0$ spots than the pattern from the $WSe_2$ region in Figure 1c, again, as expected from the relative difference in atomic numbers of the constituent elements, $\Delta Z$, and hence, the polarity, for the two materials. This difference in polarity is also reflected in Figure 4d, where the difference between the intensities the $010$ and $0\bar{1}0$ peaks for $WS_2$ is greater than that for $WSe_2$.

## 5. Conclusions

Monolayer TMD materials act as sufficiently strong phase objects that measurable asymmetry exists between the intensities of Friedel pairs in their diffraction patterns. The effect is strong enough to be useful for mapping the direction of polarity in non-centrosymmetric TMD monolayers, enabling us to rapidly map polarity domains of the crystal across orders of magnitude in length scale using diffraction-contrast imaging. We identify the origin of the asymmetry in terms of a power-series expansion of the strong phase approximation, which leads to a similar phase analysis in terms of three-phase invariants and scattering paths as that of the bulk Bloch-wave based multiple scattering analysis for bulk crystals. In using this effect for mapping polarity, it is also important to remember that for few-layer samples, especially those that lack inversion symmetry along the out-of-plane direction, sample mistilts can also lead to asymmetric contrast - although it is readily distinguishable from polarity by inspection of the EMPAD data – the second order peaks will change contrast from mistilt, but not polarity.

**Funding**

This work was supported by the National Science Foundation (NSF) through the Platform for the Accelerated Realization, Analysis, and Discovery of Interface Materials (PARADIM; DMR-1539918) (PD) and the Air Force Office of Scientific Research through the 2D Electronics MURI grant FA9550-16-1-0031 (MC). Support for the EMPAD development was provided by the U.S. Department of Energy, grant DE-FG02-10ER46693. The adaptation to the STEM was supported by the Kavli Institute at Cornell for Nanoscale Science. YH, electron microscope and facilities were supported by the Cornell Center for Materials Research, through the National Science Foundation MRSEC program, award #DMR 1719875.

**Acknowledgements**

The authors acknowledge microscopy support from John Grazul and Mariena Silvestry Ramos. We thank Kayla X. Nguyen, Mark Tate, Prafull Purohit, and Sol Gruner for help with the pixel array detector.

# References

1. Huang, P.Y., et al., *Grains and grain boundaries in single-layer graphene atomic patchwork quilts.* Nature, 2011. **469**(7330): p. 389-392.
2. Brown, L., et al., *Twinning and twisting of tri-and bilayer graphene.* Nano letters, 2012. **12**(3): p. 1609-1615.
3. Kim, C.-J., et al., *Stacking order dependent second harmonic generation and topological defects in h-BN bilayers.* Nano letters, 2013. **13**(11): p. 5660-5665.
4. Meyer, J.C., et al., *The structure of suspended graphene sheets.* Nature, 2007. **446**(7131): p. 60-63.
5. Van Der Zande, A.M., et al., *Grains and grain boundaries in highly crystalline monolayer molybdenum disulphide.* Nature materials, 2013. **12**(6): p. 554-561.
6. Splendiani, A., et al., *Emerging photoluminescence in monolayer MoS2.* Nano letters, 2010. **10**(4): p. 1271-1275.
7. Georgiou, T., et al., *Vertical field-effect transistor based on graphene-WS2 heterostructures for flexible and transparent electronics.* Nature nanotechnology, 2013. **8**(2): p. 100-103.
8. Radisavljevic, B., et al., *Single-layer MoS2 transistors.* Nature nanotechnology, 2011. **6**(3): p. 147-150.
9. Friedel, G., *Sur les symétries cristallines que peut révéler la diffraction des rayons Röntgen.* CR Acad Sci, 1913. **157**: p. 1533-1536.
10. Buerger, M., *The crystallographic symmetries determinable by X-ray diffraction.* Proceedings of the National Academy of Sciences, 1950. **36**(5): p. 324-329.
11. Bethe, H., *Theorie der beugung von elektronen an kristallen.* Annalen der Physik, 1928. **392**(17): p. 55-129.
12. Cowley, J.M. and A.F. Moodie, *The scattering of electrons by atoms and crystals. I. A new theoretical approach.* Acta Crystallographica, 1957. **10**(10): p. 609-619.
13. Howie, A.t. and M. Whelan. *Diffraction contrast of electron microscope images of crystal lattice defects. II. The development of a dynamical theory.* in *Proceedings of the Royal Society of London A: Mathematical, Physical and Engineering Sciences*. 1961. The Royal Society.
14. Takagi, S., *Dynamical theory of diffraction applicable to crystals with any kind of small distortion.* Acta Crystallographica, 1962. **15**(12): p. 1311-1312.
15. Kirkland, E.J., *Advanced computing in electron microscopy*. 2010: Springer Science & Business Media.
16. Brown, L., et al., *Twinning and Twisting of Tri- and Bilayer Graphene.* Nano Letters, 2012. **12**(3): p. 1609-1615.
17. McMullan, G., et al., *Comparison of optimal performance at 300keV of three direct electron detectors for use in low dose electron microscopy.* Ultramicroscopy, 2014. **147**: p. 156-163.
18. McGrouther, D., et al., *Use of a hybrid silicon pixel (Medipix) detector as a STEM detector.* Microscopy and Microanalysis, 2015. **21**: p. 1595.
19. Caswell, T.A., et al., *A high-speed area detector for novel imaging techniques in a scanning transmission electron microscope.* Ultramicroscopy, 2009. **109**(4): p. 304-311.
20. Faruqi, A., R. Henderson, and L. Tlustos, *Noiseless direct detection of electrons in Medipix2 for electron microscopy.* Nuclear Instruments and Methods in Physics Research Section A: Accelerators, Spectrometers, Detectors and Associated Equipment, 2005. **546**(1): p. 160-163.
21. Tate, M.W., et al., *High dynamic range pixel array detector for scanning transmission electron microscopy.* Microscopy and Microanalysis, 2016. **22**(1): p. 237-249.
22. Hirsch, P.B., A. Howie, and M.J. Whelan, *A kinematical theory of diffraction contrast of electron transmission microscope images of dislocations and other defects.* Philosophical Transactions of

　　　the Royal Society of London A: Mathematical, Physical and Engineering Sciences, 1960. **252**(1017): p. 499-529.
23. Born, M. and E. Wolf, *Principles of optics: electromagnetic theory of propagation, interference and diffraction of light*. 2013: Elsevier.
24. Moliere, G., *Theorie der streuung schneller geladener teilchen i. einzelstreuung am abgeschirmten coulomb-feld.* Zeitschrift für Naturforschung A, 1947. **2**(3): p. 133-145.
25. Brillouin, L., *La mécanique ondulatoire de Schrödinger; une méthode générale de résolution par approximations successives.* Compt. Rend. Hebd. Seances Acad. Sci., 1926. **183**: p. 24-26.
26. Kramers, H.A., *Wellenmechanik und halbzahlige Quantisierung.* Zeitschrift für Physik A Hadrons and Nuclei, 1926. **39**(10): p. 828-840.
27. Wentzel, G., *Eine verallgemeinerung der quantenbedingungen für die zwecke der wellenmechanik.* Zeitschrift für Physik A Hadrons and Nuclei, 1926. **38**(6): p. 518-529.
28. Zuo, J.M. and J.C.H. Spence, *Advanced Transmission Electron Microscopy - Imaging and Diffraction in Nanoscience*. 2017, New York: Springer 729.
29. Mannebach, E.M., et al., *Dynamic structural response and deformations of monolayer MoS2 visualized by femtosecond electron diffraction.* Nano letters, 2015. **15**(10): p. 6889-6895.

# Appendix A: Tracing the Explicit Form Factor Dependence of Polarity

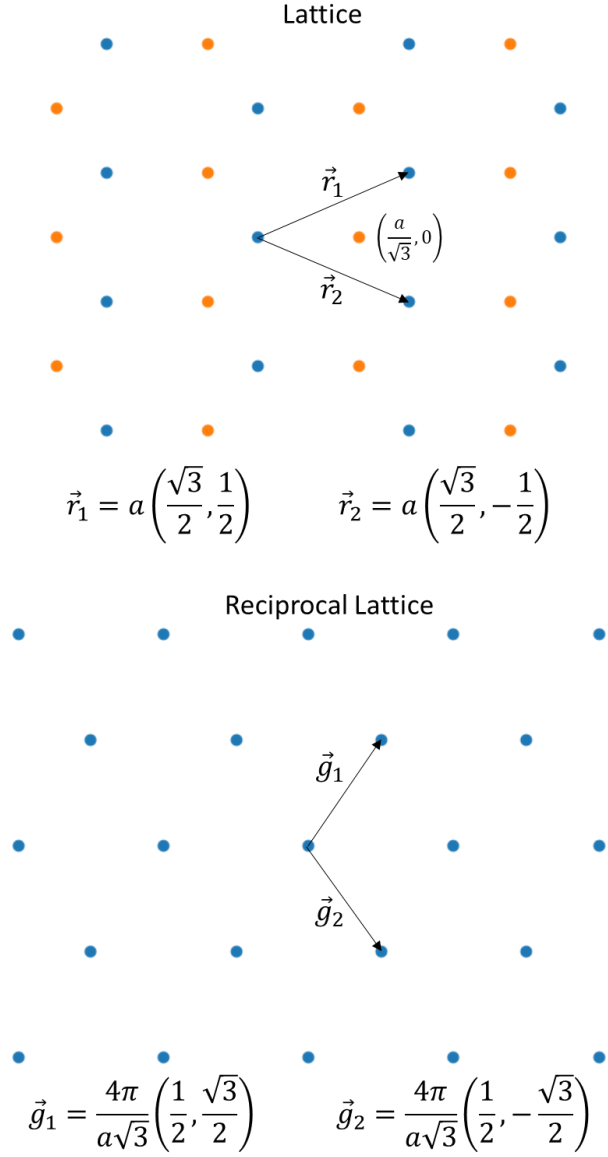

Figure A.1: Diagram of hexagonal lattice and its reciprocal lattice with basis vectors written.

Here we derive an explicit expression for the leading-order contribution to the polarity in terms of the atomic structure factors for atoms in a hexagonal lattice, a common structure for 2D materials such as graphene and BN, or 2D projections of transition metal dichalcogenides such as $MoS_2$, $WSe_2$ and the other materials considered in the main body of this paper. For the TMD materials $AB_2$, the B site form factor is twice the atomic form factor.

We start by calculating the potential distribution in Fourier space, starting with the expression for the structure factor (equation 5):

$$F(\boldsymbol{q}) = \frac{1}{2\pi e a_0} \int dr \exp(-i\boldsymbol{q} \cdot \boldsymbol{r}) V(\boldsymbol{r}). \qquad (A.1)$$

Re-writing the potential as a sum of individually displaced a-site and b-site atom potentials $v_a(\boldsymbol{r})$ and $v_b(\boldsymbol{r})$:

$$F(\boldsymbol{q}) = \frac{1}{2\pi e a_0} \int \sum_{i,j} v_a(\boldsymbol{r} - i*\boldsymbol{r_1} - j*\boldsymbol{r_2}) + v_b\left(\boldsymbol{r} - \frac{a}{\sqrt{3}}\hat{x} - i*\boldsymbol{r_1} - j*\boldsymbol{r_2}\right) \cdot \frac{dr \exp(-i\boldsymbol{q}\cdot\boldsymbol{r})}{} \qquad (A.2)$$

Basic Fourier Transform properties transform the individual potentials into their respective form factors along with a phase due to the displacement:

$$F(\boldsymbol{q}) = \frac{1}{2\pi e a_0} \sum_{i,j} \left( f_a(\boldsymbol{q}) + f_b(\boldsymbol{q}) \exp\left(-\frac{iq_x a}{\sqrt{3}}\right) \right) \exp(-i\boldsymbol{q} \cdot (i*\boldsymbol{r_1} + j*\boldsymbol{r_2})). \qquad (A.3)$$

The phase term at the end can be re-written as a delta functions centered on the reciprocal lattice. As a matter of notation, we use lower case $\boldsymbol{g}$ to refer to basis vectors, and capital $\boldsymbol{G}$ to refer to general reciprocal space vectors. We can also more explicitly write the phase contribution due to the b-site displacement:

$$F(\boldsymbol{q}) = \frac{1}{2\pi e a_0} \sum_{h,k} \left( f_a(\boldsymbol{q}) + f_b(\boldsymbol{q}) \exp\left(-\frac{2\pi i}{3}(h+k)\right) \right) \delta(\boldsymbol{q} - h\boldsymbol{g_1} - k\boldsymbol{g_2}). \qquad (A.4)$$

We use the electron form factors parametrized by Kirkland [15] for numerical simulations. Equation (A.4) can be re-written as an amplitude and phase for an explicit $\boldsymbol{G} = h\boldsymbol{g_1} + k\boldsymbol{g_2}$:

$$F(\boldsymbol{G}) = U_{\boldsymbol{G}} e^{-i\phi_{\boldsymbol{G}}} \qquad (A.5)$$

$$U_{\boldsymbol{G}} = \frac{1}{2\pi e a_0} \sqrt{f_a^2(\boldsymbol{G}) + f_b^2(\boldsymbol{G}) + 2 f_a(\boldsymbol{G}) f_b(\boldsymbol{G}) \cos\left(\frac{2\pi}{3}(h+k)\right)} \qquad (A.6)$$

$$\phi_G = \tan^{-1} \frac{f_b(G) \sin\left(\frac{2\pi}{3}(h+k)\right)}{f_a(G) + f_b(G) \cos\left(\frac{2\pi}{3}(h+k)\right)} \tag{A.7}$$

The first-order diffraction spots are at $g_1, g_2, g_1 + g_2$ and their negatives. Since atomic form factors $f(q)$ are radially symmetric, we'll use $f_{a1}$ and $f_{b1}$ to refer to their respective form factors at the first-order scattering angle.

$$f_{a1} = f_a\left(\frac{4\pi}{a\sqrt{3}}\right), f_{b1} = f_b\left(\frac{4\pi}{a\sqrt{3}}\right) \tag{A.8}$$

$$F(g_1) = F(g_2) = F(-g_1 - g_2) = \frac{1}{2\pi e a_0}\left(f_{a1} - \frac{1}{2}f_{b1} - \frac{\sqrt{3}}{2}if_{b1}\right) \tag{A.9}$$

$$F(-g_1) = F(-g_2) = F(g_1 + g_2) = \frac{1}{2\pi e a_0}\left(f_{a1} - \frac{1}{2}f_{b1} + \frac{\sqrt{3}}{2}if_{b1}\right) \tag{A.10}$$

Second-order spots are at $2g_1 + g_2, g_1 + 2g_2, g_1 - g_2$ and their negatives. These all lie at scattering angle amplitude $q = \frac{4\pi}{a}$. Notably, these all have the same form factor.

$$f_{a2} = f_a\left(\frac{4\pi}{a}\right), f_{b2} = f_b\left(\frac{4\pi}{a}\right) \tag{A.11}$$

$$F\left(|q| = \frac{4\pi}{a}\right) = \frac{1}{2\pi e a_0}(f_{a2} + f_{b2}) \tag{A.12}$$

Looking back at eq. (12) in the main text, we can evaluate the contribution of this third order term for $g_1 + g_2$. This is an example of constructive second scattering contributing to the polarity:

$$P(g_1 + g_2) = \frac{(\sigma e a_0)^3}{4\pi^3}|\Psi_0(q - g_1 - g_2)|^2$$
$$U_{g_1+g_2} U_{g_1} U_{g_2} \sin(\phi_{g_1+g_2} - \phi_{g_1} - \phi_{g_2}). \tag{A.13}$$

Looking at eq. (A.9) and (A.10), all the $U$ terms are the same, which we'll re-write as $U_1$. For the phases, $\phi_{g_1+g_2} = -\phi_{g_1} = -\phi_{g_2}$. We can re-write eq. (A.13) as:

$$P_3(\boldsymbol{g_1} + \boldsymbol{g_2}) = -\frac{(\sigma e a_0)^3}{4\pi^3}|\Psi_0(\boldsymbol{q} - \boldsymbol{g_1} - \boldsymbol{g_2})|^2 U_1^3 \sin(3\phi_{g_1}). \tag{A.14}$$

Calculating this third order term for the $-\boldsymbol{g_1} - \boldsymbol{g_2}$ spot is similar, but gives the negative result because $\phi_{-g_1-g_2} = \phi_{g_1} = -\phi_{-g_1}$:

$$P_3(-\boldsymbol{g_1} - \boldsymbol{g_2}) = \frac{(\sigma e a_0)^3}{4\pi^3}|\Psi_0(\boldsymbol{q} + \boldsymbol{g_1} + \boldsymbol{g_2})|^2 U_1^3 \sin(3\phi_{g_1}). \tag{A.15}$$

These results are consistent with Fig. 2b. Furthermore, we can rewrite the $\sin 3\phi$ and multiply with the $U$ terms to re-write eq. (A.14) and (A.15) in terms of $f_{a1}$ and $f_{b1}$:

$$\sin(3\phi_{g_1}) = 3\sin\phi \cos^2\phi - \sin^3\phi. \tag{A.16}$$

Noting that the magnitude $U_1$ times the cosine or sine gives the real and imaginary portions of the form factor, respectively:

$$U_1 \sin\phi = \frac{-f_{b1}\sqrt{3}}{4\pi e a_0}, \tag{A.17}$$

$$U_1 \cos\phi = \frac{f_{a1} - \frac{1}{2}f_{b1}}{2\pi e a_0}, \tag{A.18}$$

$$P_3(\boldsymbol{g_1} + \boldsymbol{g_2}) = \frac{-\sigma^3}{32\pi^6}|\Psi_0(\boldsymbol{q} + \boldsymbol{g_1} + \boldsymbol{g_2})|^2$$
$$* 3\left(-\frac{\sqrt{3}}{2}f_{b1}\right)\left(f_{a1} - \frac{1}{2}f_{b1}\right)^2 - \left(-\frac{\sqrt{3}}{2}f_{b1}\right)^3, \tag{A.19}$$

$$P_3(\boldsymbol{g_1} + \boldsymbol{g_2}) = \frac{3\sigma^3\sqrt{3}}{64\pi^6}|\Psi_0(\boldsymbol{q} - \boldsymbol{g_1} - \boldsymbol{g_2})|^2 f_{a1}f_{b1}(f_{a1} - f_{b1}), \tag{A.20}$$

$$P_3(-\boldsymbol{g_1} - \boldsymbol{g_2}) = -\frac{3\sigma^3\sqrt{3}}{64\pi^6}|\Psi_0(\boldsymbol{q} + \boldsymbol{g_1} + \boldsymbol{g_2})|^2 f_{a1}f_{b1}(f_{a1} - f_{b1}). \quad (A.21)$$

Now the polarity strength is explicitly dependent on the difference between form factors. Notably, this clearly shows that for materials like graphene where $f_{a1} = f_{b1}$, there is no polarity. Below is a graph of $|f_{a1} - f_{b1}|$ as a function of $|\Delta Z|$.

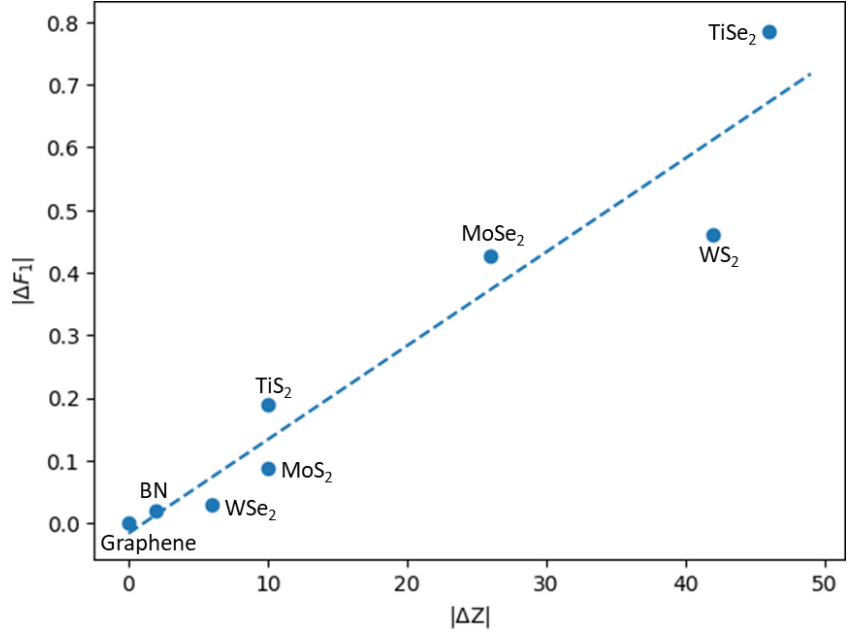

Figure A.2: Graph of difference in form factor $|f_{a1} - f_{b1}|$ as a function of difference in Z-number $|\Delta Z|$ between A and B sites for a variety of materials.

We can do a similar calculation for the second order spots at $\boldsymbol{g_1} - \boldsymbol{g_2}$ and $\boldsymbol{g_2} - \boldsymbol{g_1}$. This time, these paths are destructive since $\phi_{g_1} = -\phi_{-g_2}$. Further all the second order spots have zero phase. This means, we can easily state:

$$P(\boldsymbol{g_1} - \boldsymbol{g_2}) = 0 \quad (A.22)$$

We emphasize that this does not mean that the diffraction spot itself has no intensity, but that there is no polarity-contributing term for this spot. This is consistent with graphical construction in Fig. 2c.

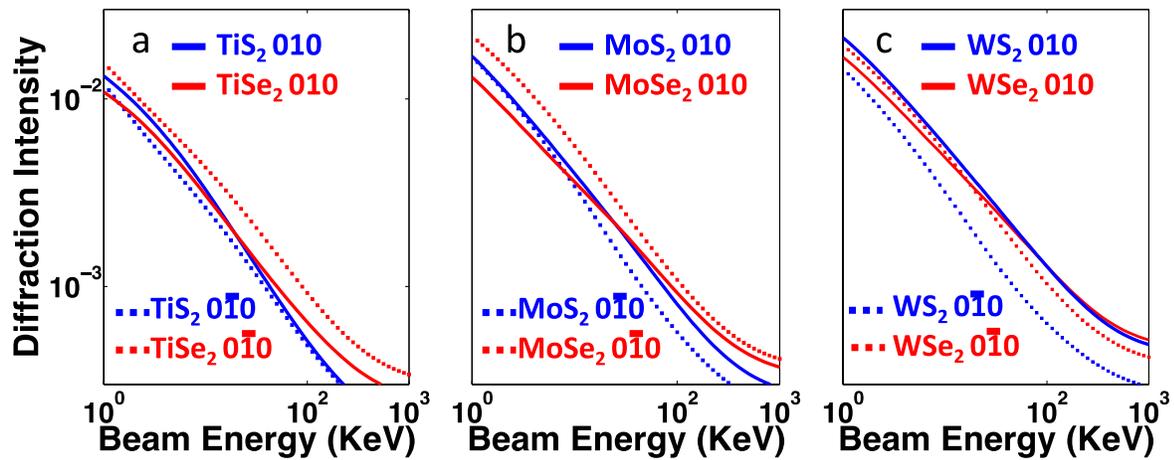

**Supplementary Figure 1. a-c,** The diffraction intensities for the $010$ and $0\bar{1}0$ peaks are plotted as a function of beam energy for the sulfides and selenides of transition metals in increasing order of atomic number, Ti, Mo and W in **a, b** and **c** respectively. The difference between the peaks is correlated with ΔZ, where ΔZ =Z(A)-n Z(B), and Z is the atomic number of A and B in the polar material $AB_n$. In **a** and **b,** we see a greater difference between the two peaks for selenides (red) than sulfides (blue), because the magnitude of ΔZ for $TiS_2$ and ΔZ for $MoS_2$ is less than that for $TiSe_2$ and $MoSe_2$ respectively. However, in **c,** the difference between the two peaks is greater for the sulfide (blue) than for the selenide (red), since the magnitude of ΔZ for $WS_2$ is greater than that for $WSe_2$.